\documentclass[3p,times,twocolumn]{elsarticle}

\usepackage{ecrc}

\volume{00}

\firstpage{1}

\journalname{Nuclear Physics B Proceedings Supplement}

\runauth{}

\jid{nuphbp}

\jnltitlelogo{Nuclear Physics B Proceedings Supplement}

\usepackage{amssymb}
\usepackage{hyperref}

\usepackage[figuresright]{rotating}

\newcommand{\alphas}{\alpha_{\rm s}}
\newcommand{\alphasmZ}{\alphas(\rm m^2_{_{\rm Z}})}
\newcommand{\sqrts}{\sqrt{\rm s}}

\newcommand{\lqcd}{\Lambda_{_{\rm QCD}}}

\newcommand{\pp}{p-p}
\newcommand{\ppbar}{{p-$\rm \bar{p}$}}
\newcommand{\epem}{e^+e^-}
\newcommand{\meff}{m{_{\rm eff}}}
\newcommand{\ximax}{\xi{_{\rm max}}}

\begin{document}

\begin{frontmatter}



\dochead{}

\title{Uncertainties on the determination of the strong coupling $\alphas$\\ from the energy evolution 
of jet fragmentation functions at low $z$} 


\author{David d'Enterria$^1$ and Redamy P\'erez-Ramos$^2$} 

\address{$^1$  CERN, PH Department, CH-1211 Geneva 23, Switzerland\\
$^2$ Department of Physics, Univ. of Jyv\"askyl\"a, P.O. Box 35, F-40014 Jyv\"askyl\"a, Finland}

\begin{abstract}
The QCD coupling $\alphas$ is determined at NLO*+NMLLA accuracy from the comparison of experimental jet data to 
theoretical predictions of the energy-evolution of the parton-to-hadron fragmentation function moments
(multiplicity, peak, width, skewness) at low fractional hadron momentum $z$. 
From the existing $\epem$ and 
$e^\pm$p jet data, we obtain $\alphasmZ$~=~0.1195$\pm$0.0021(exp)$^{+0.0015}_{-0.0}$(scale) at the Z mass. 
The uncertainties of the extracted $\alphas$ value are discussed.  
\end{abstract}


\end{frontmatter}


\section{Introduction}
\label{sec:intro}

In the chiral (massless quark) limit, the theory of the strong interaction --quantum chromodynamics (QCD)--
has a single fundamental parameter: its coupling $\alphas$ which decreases logarithmically with increasing
energy scale Q, i.e. $\alphas\propto\ln(\rm Q^2/\lqcd^2)$, starting from a value $\lqcd\!\!\approx\,$0.2~GeV where the
perturbatively-defined coupling diverges.
The running coupling $\alphas$ enters in the perturbative expansion of theoretical cross section for any hard process
involving quarks and gluons, and the uncertainty in its value is a key (sometimes dominant) component of the
theoretical error in all perturbative QCD (pQCD) predictions for any collision process involving hadrons.
Many precision fits of the Standard Model as well as searches for new physics depend on $\alphas$,
whose value needs, thus, to be determined with good accuracy and precision.\\

The determination of $\alphas$ relies on the comparison of theoretical predictions for various observables, 
obtained perturbatively at a given level of accuracy (next-to-next-to-leading order, NNLO, in most cases),  
with the corresponding experimental measurements. The extracted $\alphas$ values at
different energy scales are then compared to each other by translating the result in terms of the value at the
Z mass pole: $\alphasmZ$. The current value of $\alphasmZ$~=~0.1185$\pm$0.0006, has been obtained from a
combination of measurements at $\epem$, deep-inelastic scattering (DIS) e,$\nu$-p, and hadron-hadron colliders~\cite{PDG}: 
\begin{description}
\item (i) hadronic $\tau$ decays (N$^3$LO), (ii) hadronic W,Z decays (N$^3$LO), 
(iii) radiative heavy-quarkonia decays (NLO), and (iv) event shapes and jet rates (NNLO), in $\epem$ collisions; 
\item (v) scaling violations in parton distribution functions (NNLO), and jet cross sections (NLO), in e,$\nu$-p DIS;
\item (vi) jet cross sections and angular correlations (NLO), and top-quark cross sections (NNLO), in \pp,\ppbar\ collisions.
\end{description}
In addition, comparisons of the predictions for different short-distance observables computed through lattice
and pQCD methods provide extra ``data points'' for the determination of the $\alphasmZ$ world-average,
although the low systematic uncertainty assigned to such an approach has been questioned~\cite{Altarelli:2013bpa}.
The current $\alphas$ uncertainty is of order $\pm$0.5\%, although a more conservative estimate
sets it at the $\pm$1\% level~\cite{Altarelli:2013bpa}, making of $\alphas$ the least precisely known of all
fundamental couplings in nature. In this context, having at hand extra independent approaches to determine $\alphas$ 
--with experimental and theoretical uncertainties different than those of the methods currently
used-- would be an obvious asset.\\

Attempts were carried out in the past (see e.g.~\cite{Akrawy:1990ha}) to try to measure $\lqcd$ (or,
equivalently, $\alphas$) using parton-to-hadron fragmentation functions (FFs) in the region of low fractional momenta 
$z=p_{\rm h}/p_{\rm parton}$ via the Modified Leading Logarithmic Approximation (MLLA)~\cite{mlla} 
which resums the soft and collinear singularities present in this region of phase-space. However, these older
calculations were limited to LO 
approximations with a number of simplifying assumptions: (i) ad hoc cuts in the experimental distributions, (ii) simple fits
in a restricted FF range, (iii) number of quark-flavours fixed to $N_f$~=~3, and (iv) use of only one or two FF
moments (Gaussian approximation). As a result, very imprecise (or even inconsistent) values of $\lqcd\approx$~80--600~MeV
were obtained.  In Ref.~\cite{NMLLA_NLO} we have presented a novel extraction of $\alphas$ from the energy
evolution of the first four moments of the parton-to-hadron FFs including, for the first
time, resummations of higher-order (NNLL or NMLLA) logarithms and NLO running-coupling corrections. The approach has been
successfully tested with the world experimental jet data from $\epem$ annihilation~\cite{NMLLA_NLO} and
deep-inelastic $e^\pm$p collisions~\cite{moriondqcd14}. We provide here a more detailed assessment of the
systematic uncertainties involved in our $\alphas$ extraction.

\section{NLO*+NMLLA parton-to-hadron fragmentation functions}
\label{sec:FFs}

\begin{figure}[htpb!]
\centerline{
\includegraphics[width=0.99\linewidth]{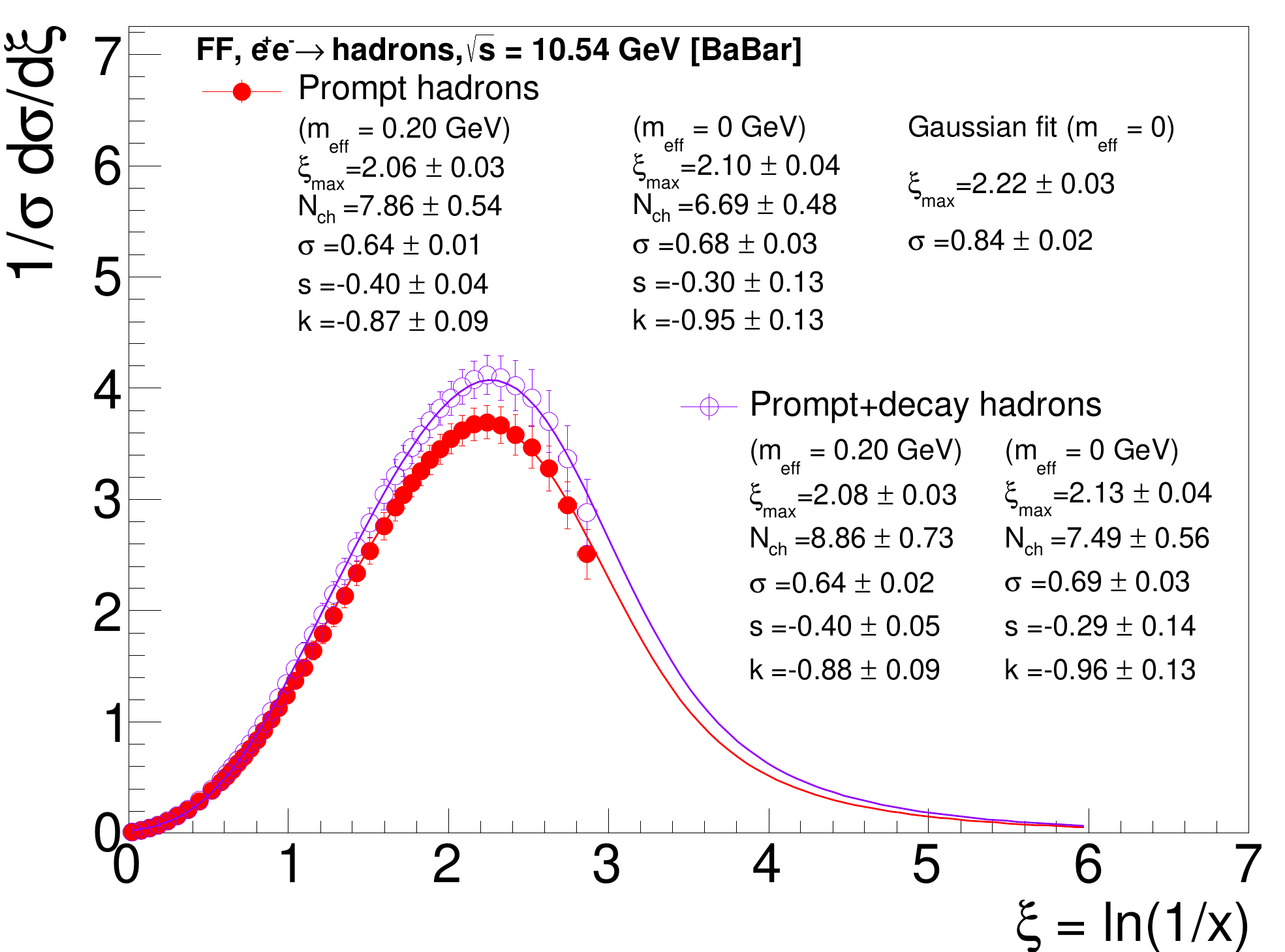}
}
\caption[]{
Jet fragmentation function measured by BaBar for prompt (closed dots) and inclusive (open dots)
charged-hadrons in $\epem$ at $\sqrts=$~10.54~GeV~\cite{babar} fitted to the DG Eq.~(\ref{eq:DG}).
The obtained moments are listed for two choices of the effective mass ($\meff$~=~0,0.2~GeV). 
}
\label{fig:FFbabar}
\end{figure}

The parton-to-hadron FF, $D_{\rm i\to h}(z,\rm Q)$, encodes the probability that parton $i$ fragments into a
hadron $h$ carrying a fraction $z$ of the parent parton's momentum. 
Due to colour coherence and gluon-radiation interference inside a parton shower, not the softest partons but
those with intermediate energies ($E_h\propto E_{\rm jet}^{0.3}$) multiply most effectively in QCD cascades, 
leading to a single-particle spectrum with a typical ``hump-backed plateau'' (HBP) shape as a function of
$\xi=\ln(1/z)$ (Fig.~\ref{fig:FFbabar}). Such an HBP shape can be parametrized, without any loss of generality,
as a distorted Gaussian (DG) which depends on the original energy of the parton, 
$Y = \ln{\rm E\theta/Q_{_{0}}}$, evolved down to a shower cut-off scale $\lambda = \ln(\rm Q_{_{0}}/\lqcd)$:
\begin{equation}
\hspace{-0.7cm}D(\xi,Y,\lambda) = {\cal N}/(\sigma\sqrt{2\pi})\cdot e^{\left[\frac18k-\frac12s\delta-
\frac14(2+k)\delta^2+\frac16s\delta^3+\frac1{24}k\delta^4\right]}\,, 
\label{eq:DG}
\end{equation}
where $\delta=(\xi-\bar\xi)/\sigma$, with moments:
${\cal N}$ (hadron multiplicity inside the jet), $\bar\xi$ (DG peak position), 
$\sigma$ (DG width), $s$ (DG skewness), and $k$ (DG kurtosis). Fig.~\ref{fig:FFbabar} shows a typical
example of FFs fitted to Eq.~(\ref{eq:DG}) with the corresponding extracted moments.
Since the measured final-state hadrons are massive and the theoretical calculations assume
massless partons/hadrons (i.e. $\xi_{\rm p}=\xi_{\rm E}$), the  
expression (\ref{eq:DG}) for $\xi$ needs to be modified by introducing an effective mass, via 
$E=\sqrt{p^2+\meff^2}$, with the corresponding Jacobian determinant correction. 
Such a change of variable only affects the DG fit for the softest hadrons (i.e. for 
the highest $\xi$ values)~\cite{NMLLA_NLO}.

The process of parton radiation and splitting occurring in a jet shower,
i.e. the evolution of the FF from a scale Q to Q' before the final parton-to-hadron transition, is governed by
the DGLAP~\cite{dglap} equations (at large $z\gtrsim 0.1$, i.e. low $\xi\lesssim$~2) and by MLLA approaches
(at small $z$, i.e. large $\xi\gtrsim$~2).
In Ref.~\cite{NMLLA_NLO} we have solved the set of integro-differential
equations for the FF evolution including\footnote{'NLO*' indicates 
missing NLO terms in the splitting functions.} NLO* $\alphas$ and next-to-MLLA corrections,
by expressing the Mellin-transformed hadron distribution in terms of the $\gamma$ anomalous dimension: 
$D\simeq C(\alphas(t))\exp\left[\int^t \gamma(\alphas(t')) dt\right]$ for $t=\ln \rm Q$, resulting in an
expansion in half-powers of $\alphas$: 
$\gamma\sim {\cal O}_{_{\rm DLA}}(\sqrt{\alphas})+{\cal O}_{_{\rm MLLA}}(\alphas)+{\cal O}_{_{\rm
 NMLLA}}(\alphas^{^{3/2}})+\cdots$, of which two new higher-order terms have been computed for the first time.
The corresponding NLO*+NMLLA dependencies
of the 5 moments of Eq.~(\ref{eq:DG}) as a function of $Y$ and $\lambda$,
have been derived. 
\begin{figure*}[htpb!]
\centerline{
\includegraphics[width=0.505\linewidth]{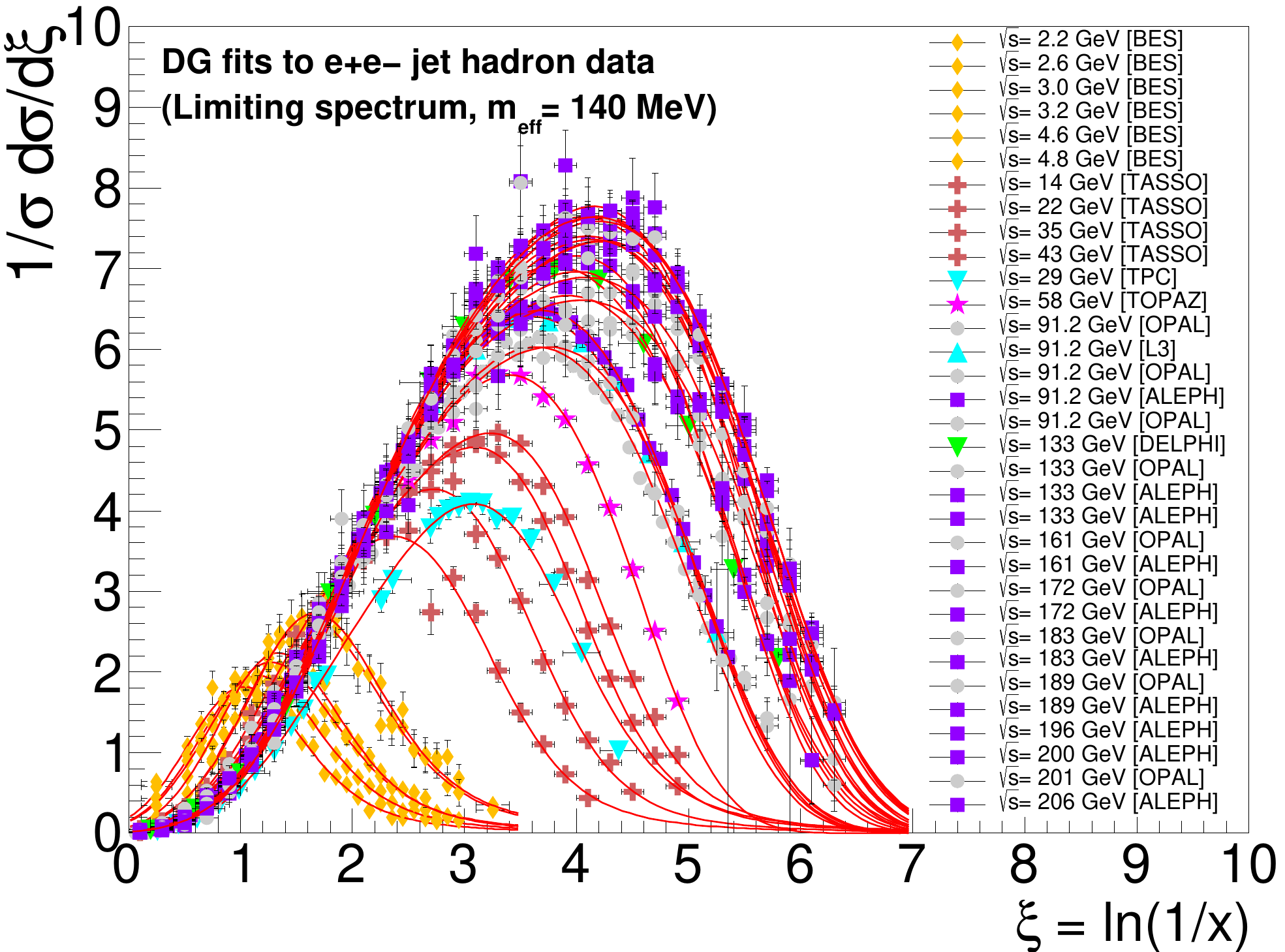}
\includegraphics[width=0.495\linewidth]{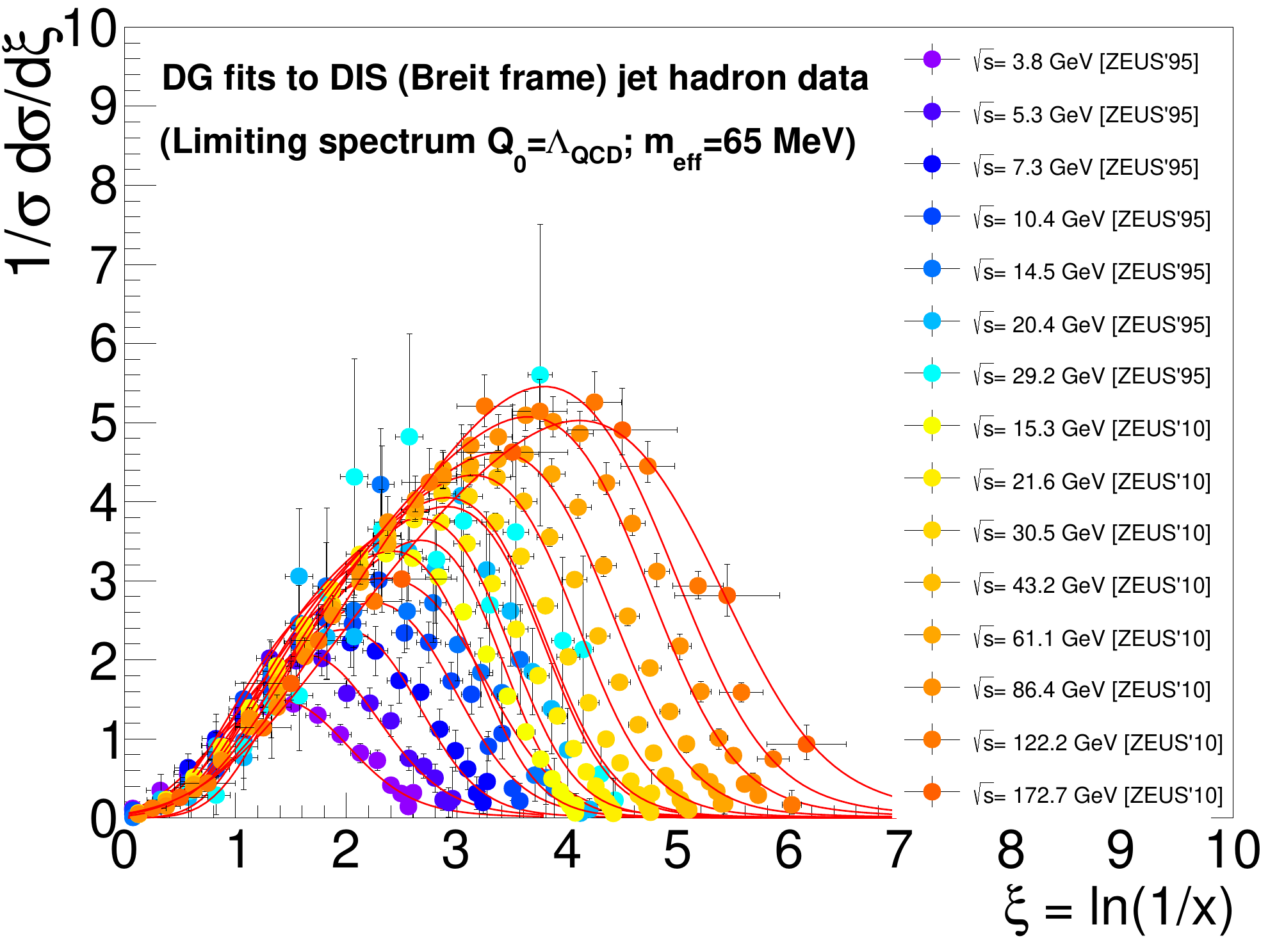}
}
\caption[]{
Charged-hadron distributions in jets as a function of $\xi=\ln(1/z)$ measured 
in $\epem$ at $\sqrts\approx$~2--200~GeV~\cite{NMLLA_NLO} (left) and
$e^\pm$p (Breit frame) at $\sqrts\approx$~4--180~GeV~\cite{moriondqcd14} (right),
individually fitted to the distorted Gaussian, Eq.~(\ref{eq:DG}), with the hadron mass corrections ($\meff$~=~140,\,65~MeV) quoted.
}
\label{fig:DGfits}
\end{figure*}
Evolving the fragmentation functions down to the lowest possible
scale, Q$_{_{0}} \to \lqcd$ (i.e. $\lambda = 0$, or ``limiting spectrum''), one obtains expressions for the
HBP moments which depend {\it only} on $\lqcd$. 
Such an approach is justified for
infrared-safe observables by the ``local parton-hadron duality'' hypothesis 
which states that the distribution of partons in jets is identical to that of the final hadrons (up to a
constant ${\cal K}_{\rm ch}$ which only affects the absolute normalization of the HBP distribution). 
Thus, by fitting  the experimental hadron jet data at various energies to the DG parametrization
(\ref{eq:DG}), one can determine $\alphas$ from the corresponding energy-dependence of its fitted moments.
Figure~\ref{fig:DGfits} shows the DG fits of the FFs measured in $\epem$ (32 datasets, left plot)~\cite{NMLLA_NLO} 
and DIS (15 datasets, right plot)~\cite{moriondqcd14}. We note that the DIS FFs need to be scaled up by a
factor of 2 since only half of the total hadron production (in the current hemisphere of the Breit frame)
is measured. For increasing jet energies the FF peak shifts to higher $\xi$, and its width broadens.

\section{Experimental uncertainties on $\alphas$}
\label{sec:exp_uncert}

The extraction of the FF moments in the DG fit has two possible sources of uncertainty: 
(i) the choice of the $\meff$ factor to account for finite hadron-mass corrections, and 
(ii) the use of experimental FFs with slightly different final charged-hadron definitions
(the different experiments require tracks coming from within 1--10~cm from the primary interaction vertex,
including or not a fraction of secondary hadrons from weak K$_s^0$ and $\Lambda$ decays).
To estimate both effects we have taken the recent high-quality prompt and inclusive pion/kaon/proton FFs 
measured by the BaBar experiment~\cite{babar}, and fitted their total (summed) distribution to DGs
with varying values of $\meff$. The resulting fitted moments for $\meff$~=~0--0.2~GeV are listed in Fig.~\ref{fig:FFbabar}. 
In addition, since some older FF studies extracted the peak position and width of the single-hadron
distributions assuming a pure Gaussian shape, we list also the extracted values of $\ximax$ and $\sigma$ in
the Gaussian approximation in the top-right legend of Fig.~\ref{fig:FFbabar}. The main findings are as follows:
\begin{description}
\item (i) The fit uncertainties of 
all the extracted moments cover well the considered range of $\meff\!\!=$0--0.2~GeV variations.
The best fits (minimum $\chi^2$) of the experimental FFs to the DG (Fig.~\ref{fig:DGfits}) are obtained for
$\meff\,$=~130$\pm$30~MeV ($\epem$) and 70$\pm$20~MeV (DIS).
\item (ii) The DG moments obtained for the prompt and inclusive charged-hadrons FFs are all consistent
within their associated uncertainties except, as expected, for the total multiplicity N$_{\rm ch}$
which is a factor of 8--10\% smaller for the primary-hadrons FFs.
\item (iii) The obtained FF maximum $\ximax$ and width $\sigma$ assuming a Gaussian distribution
  around the peak, overestimate their values by 6\% and 20\% respectively compared to the more realistic DG fits.
\end{description}
The conclusion is that our FF analysis is robust with respect to hadronization corrections,
and that only the extracted value of N$_{\rm ch}$ needs to be reduced by a factor of (9$\pm$1)\% if the
experimental distributions have not been corrected for secondary weak decays. In addition, the FF peaks and
widths extracted in Gaussian fits in the oldest data analyses overestimate these parameters by 6\% and 20\% 
respectively compared to the DG fits.\\

\begin{figure*}[htpb!]
\begin{minipage}{0.99\linewidth}
\includegraphics[width=0.90\linewidth,height=10.5cm]{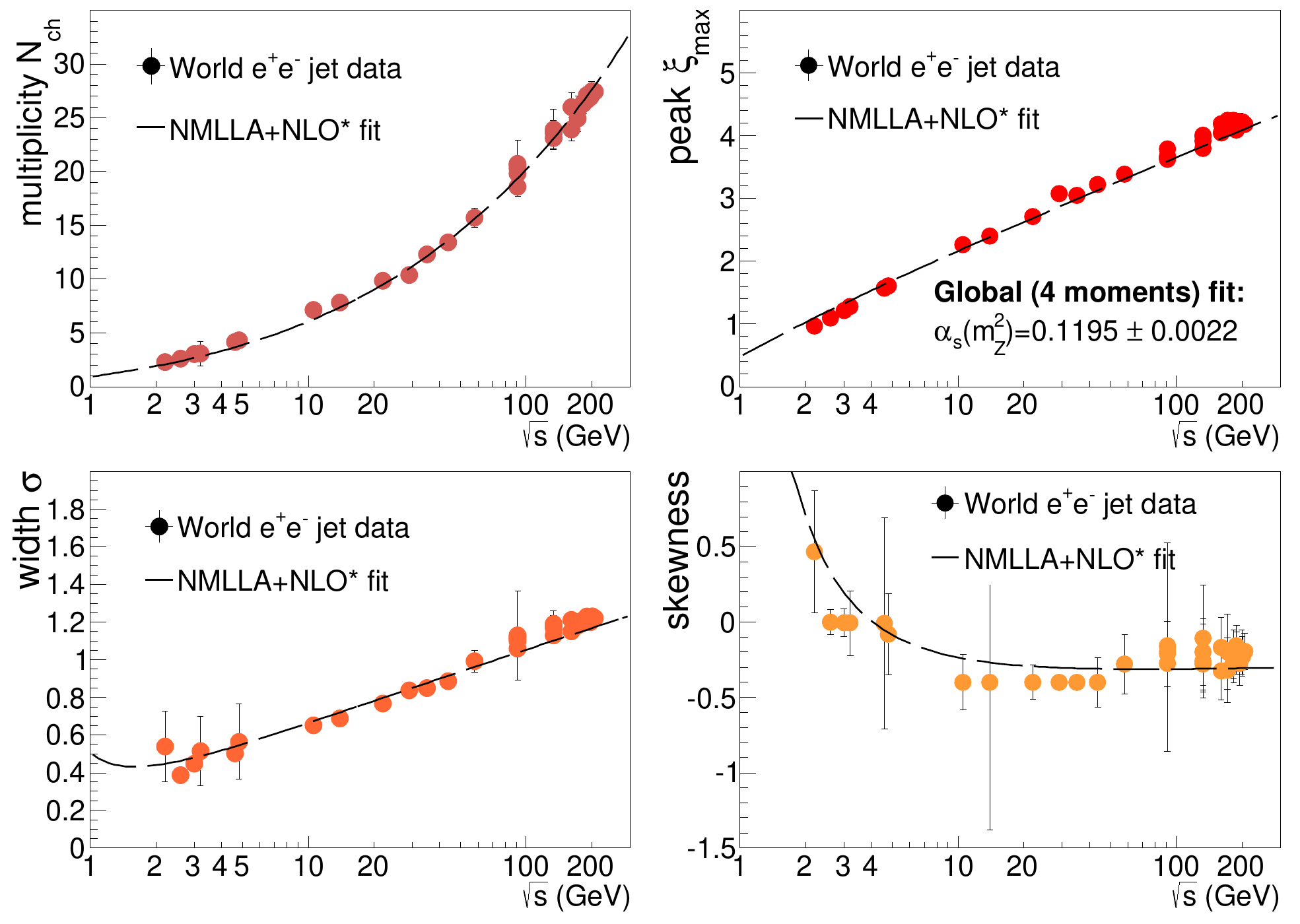}
\vspace{4mm}
\end{minipage}
\begin{minipage}{0.99\linewidth}
\includegraphics[width=0.90\linewidth,height=10.5cm]{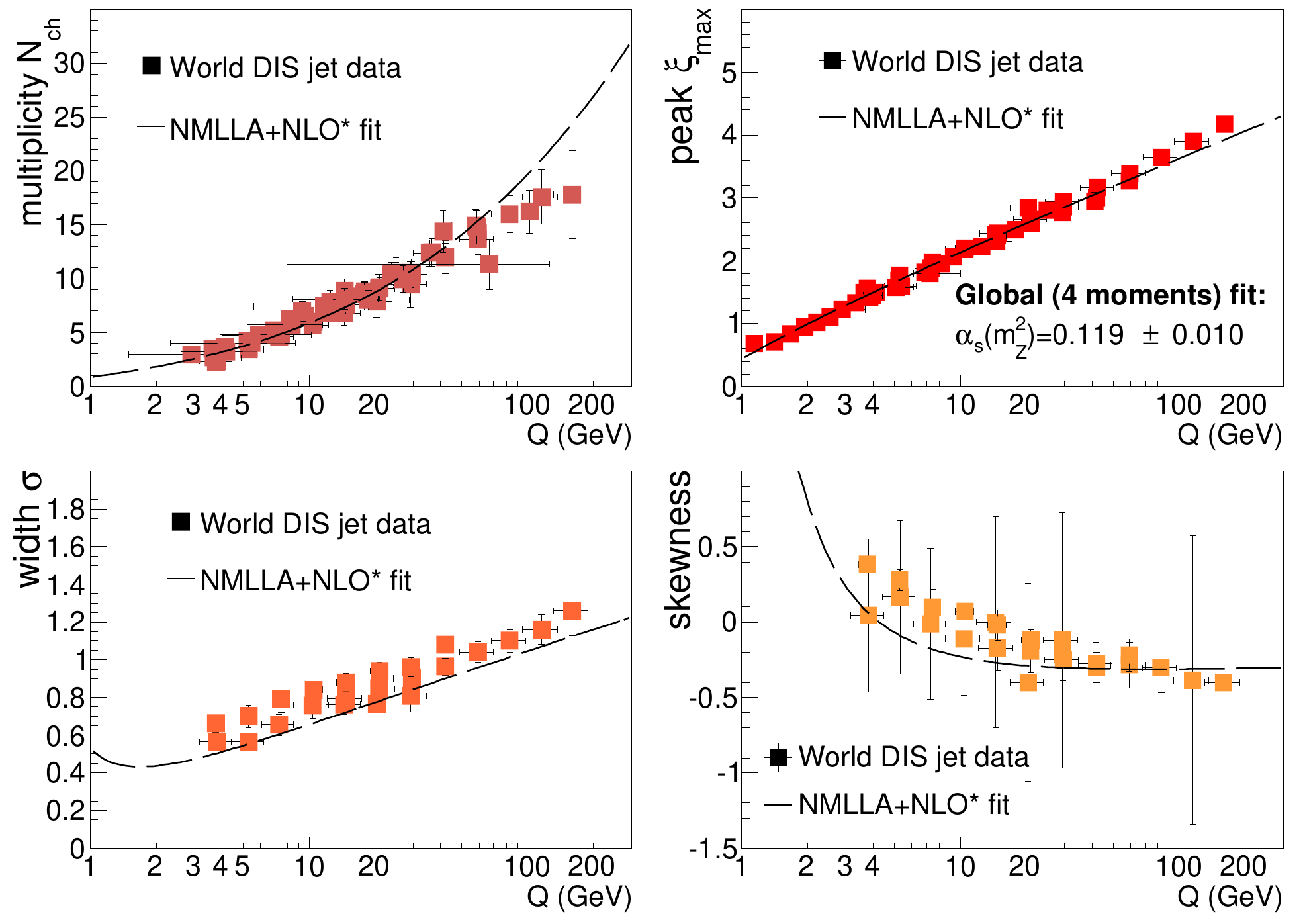}
\end{minipage}
\caption[]{Energy evolution of the moments (charged hadron multiplicity, peak, width and skewness) 
of the jet FFs measured in $\epem$ collisions at $\sqrts\approx$~2--200~GeV~\cite{NMLLA_NLO} (top panels) and 
$e^\pm$p collisions (Breit frame) at $\sqrts\approx$~4--180~GeV~\cite{moriondqcd14} (bottom panels).
The extracted values of the 
NLO $\alphasmZ$ are quoted for the the two independent data-sets fits.} 
\label{fig:3}
\end{figure*}

From the parton-to-hadron FF moments obtained for jets of various energies, one can extract $\alphas$ by
fitting their combined evolutions to our NLO*+NMLLA predictions, which include also corrections to account for
the increasing number of flavours, $N_f$~=~3,4,5, at the corresponding heavy-quark production thresholds:
E$_{\rm jet}>$~m$_{\rm charm,bottom}\approx$~1.3,4.2~GeV. 
The energy dependence of the multiplicity, peak, width and skewness of the jet FFs are shown
in Fig.~\ref{fig:3} for $\epem$ annihilation at $\sqrts\approx$~2--200~GeV~\cite{NMLLA_NLO} (top panels),
and for DIS $e^\pm$p and $\nu\,$p collisions in the range of four-momentum transfers
${\cal Q}\approx$~4--180~GeV~\cite{moriondqcd14} (lower panels). We note that the DIS plot includes more 
moments (published in previous studies) than those obtained directly fitting the data of Fig.~\ref{fig:DGfits}, right.
The hadron multiplicities measured in DIS jets (scaled by a factor of two to cover both hemispheres) are
smaller (especially at high energy) than those measured in $\epem$ collisions~\cite{NMLLA_NLO}, a 
fact pointing likely to limitations in the FF measurement only in half (current Breit) $e^\pm$p hemisphere
and/or in the determination of the relevant ${\cal Q}$ kinematic variable.
The DIS HBP widths show also larger scatter than observed in $\epem$, which is not unexpected as we are
including moments fitted in the Gaussian approximation which overestimates their value by about 20\% as
discussed in the previous section. The skewness 
has the largest uncertainties of all moments, and the kurtosis (not shown) is very close to zero and not
properly reproduced by the calculations~\cite{NMLLA_NLO}. The most robust FF moment for the determination
of $\lqcd$ is the peak position $\xi_{\rm max}$ which has the simplest theoretical expression for its
NLO*+NMLLA energy evolution~\cite{NMLLA_NLO}, and it is largely insensitive to the uncertainties
(finite-mass corrections, secondary decays, DG fit) associated with our method.\\ 

The fits of the energy-dependencies of the HBP moments allow one to extract QCD coupling values:
$\alphasmZ$~=~0.1195~$\pm$~0.0022 (from the $\epem$ data)~\cite{NMLLA_NLO}, and
$\alphasmZ$~=~0.119~$\pm$~0.010 (from DIS jets alone)~\cite{moriondqcd14}, in 
perfect agreement with the current world average. The quoted propagated $\alphasmZ$ uncertainties 
have been obtained following the ``$\chi^2$ averaging'' method~\cite{PDG}. Namely, we fit first 
the energy-dependence for each individual moment to its corresponding theoretical prediction and if the
goodness-of-fit $\chi^2$ is larger than the number of degrees of freedom (ndof), all the data-points of
the moment are enlarged by a common factor such that $\chi^2$/ndof equals unity. We then perform the
combined fit of {\it all} the energy-evolutions of the four moments letting $\lqcd$ as the {\it single}
free parameter. Such an approach takes into account in a well defined manner any correlations among the four
extractions of $\alphas$, as well as any missing extra systematic uncertainties. Our discussion of the
experimental uncertainties indicates that such an error assignment covers perfectly well the range of $\alphas$
variations induced by finite hadron-mass corrections, experimental final-hadron definitions, and DG fits.

\section{Theoretical uncertainties on $\alphas$}
\label{sec:th_uncert}

The advantages of our $\alphas$ determination compared to all other methods used so far 
are two-fold:
\begin{description}
\item (i) Our approach relies on a theoretically convergent resummation of soft and
collinear logs down to Q$_{_{0}} \to \lqcd$ 
and, thus, it is much more robust with
respect to corrections due to the transition from partons to hadrons. Hadronization uncertainties are
(much) smaller and limited to the choice of the $\meff$ factor accounting for finite-hadron mass 
effects as discussed in the previous Section.
\item (ii) Our expansion in half-powers of $\alphas$ includes almost the full set of $\alphas$ NLO corrections
plus NNLL terms. The latter log corrections partially account for contributions from missing (NNLO) terms and,
in this sense, our approach is more complete than existing $\alphas$ extractions at NLO accuracy. 
\end{description}

\begin{figure}[htpb!]
\includegraphics[width=0.99\linewidth]{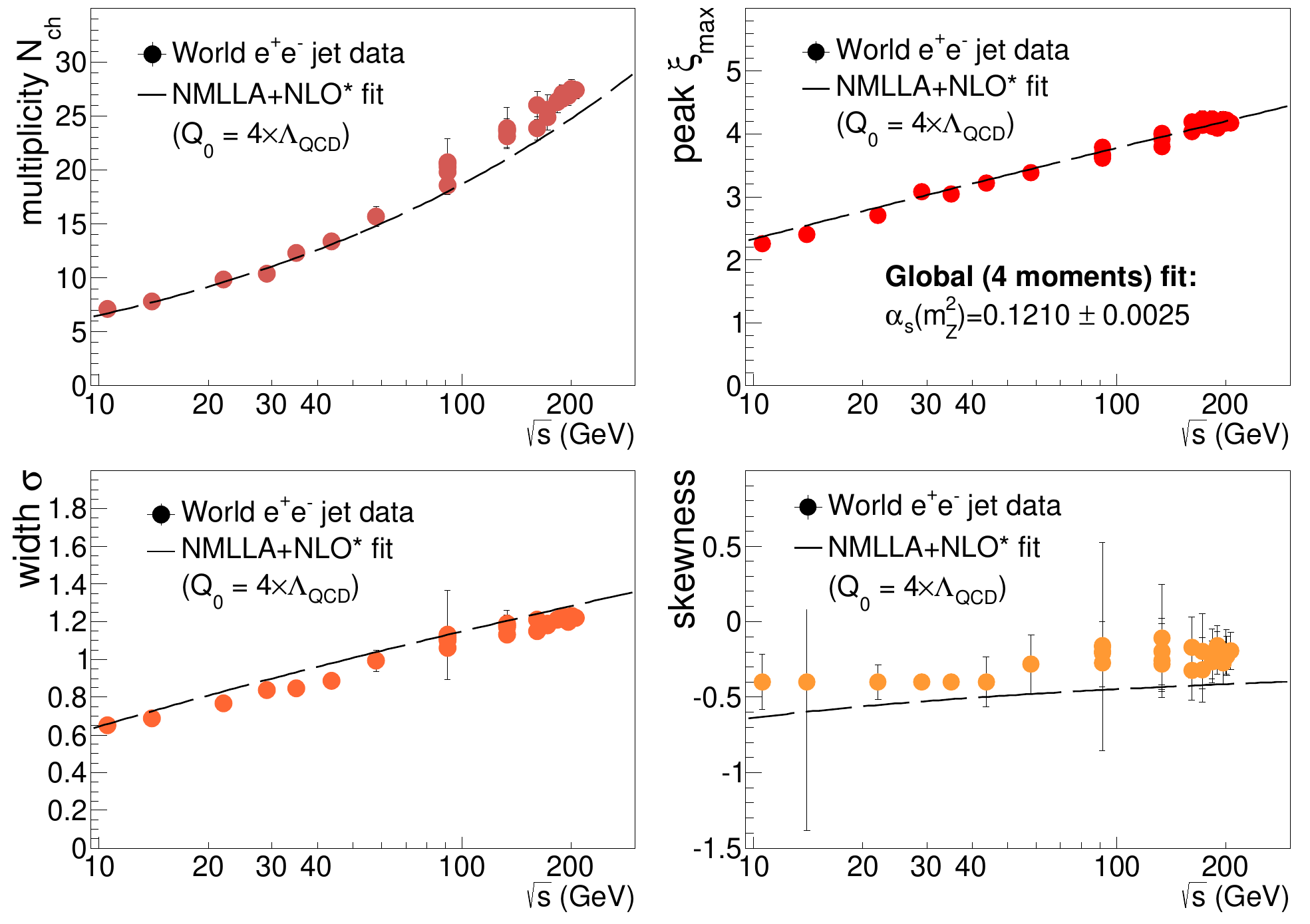}
\caption[]{Energy evolution of the moments (charged hadron multiplicity, peak, width and skewness) 
of the jet FFs measured in $\epem$ collisions at $\sqrts\approx$~10--200~GeV, fitted to the NLO*+NMLLA
predictions evaluated at a scale $\lambda$~=~1.4 (i.e. Q$_{_{0}} = 4\,\lqcd\approx$~1~GeV).}
\label{fig:4}
\end{figure}

Notwithstanding this last item, the state-of-the-art $\alphas$ determinations are at NNLO accuracy~\cite{PDG}
whereas our method is NLO*+NNLL. In order to estimate the size of theoretical uncertainties associated with
missing fixed-order terms in our truncation of the $\alphas$ expansion, we have redone the fit of the 
moments versus jet energy for a different value of the scale $\lambda = \ln(\rm Q_{_{0}}/\lqcd)$ at which the
parton evolution is stopped. 
We vary $\lambda$ from the default limiting-spectrum value ($\lambda$~=~0, i.e. Q$_{_{0}} =\lqcd$) 
to $\lambda$~=~1.4 (i.e. Q$_{_{0}} = 4\,\lqcd\approx$~1~GeV)
and fit the experimental data for jets energies in the range Q~=~10--200~GeV, so as to leave some room for
parton QCD evolution, i.e. avoiding jet data with energies too close to the shower cutoff at 1~GeV. In the relevant range of
experimental jet energies $E_{\rm jet\, max}\approx$~2,10--200~GeV, the average scale at which $\alphas$ is
effectively evaluated in our approach is given by the geometric mean: Q~=~$\sqrt{E_{\rm jet\, max}\times \rm
  Q_{_{0}}}\approx$~0.6--2~GeV (for the limiting spectrum) and Q~=~3--14~GeV (for the $\lambda$~=~1.4
case). The results of the fits for the $\epem$ data are shown in Fig.~\ref{fig:4}. We obtain
$\alphasmZ$~=~0.1210$\,\pm\,$0.0025, which is consistent with the value determined in the limiting spectrum 
case, $\alphasmZ$~=~0.1195$\,\pm\,$0.0022, although larger by +0.0015. We assign this difference as a (positive)
source of systematic error associated with the scale uncertainty of our calculations.

\section{Conclusions}
\label{sec:conclusion}

\begin{figure}[htpb!]
\centerline{
\includegraphics[width=0.99\linewidth]{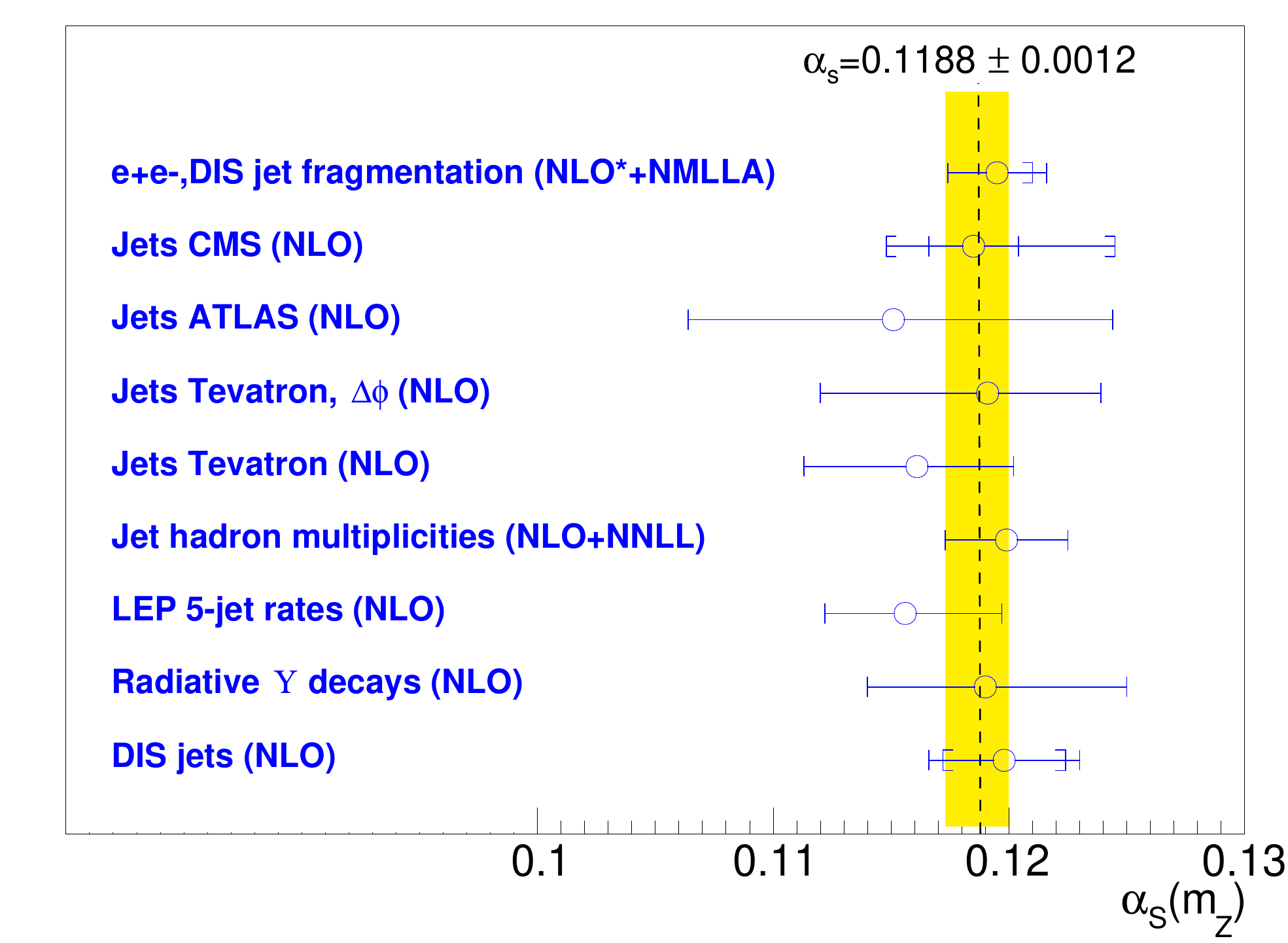}
}
\caption[]{Summary of NLO $\alphas$ determinations from different methods. 
The error ``brackets'' (if present) indicate the theoretical uncertainties of each extraction.
The dashed line and shaded (yellow) band indicate their weighted average (listed also on the top).
}
\label{fig:alphas_NLO}
\end{figure}

The QCD coupling has been determined at NLO*+NMLLA accuracy from the energy evolution of the moments of
the parton-to-hadron fragmentation functions measured in the E$_{\rm jet}=$~2--200~GeV range in $\epem$
and DIS collisions: $\alphasmZ$~=~0.1195~$\pm$~0.0022(fit), and $\alphasmZ$~=~0.119~$\pm$~0.010(fit), respectively. 
We have studied in detail the experimental and systematic uncertainties associated with our
$\alphas$ extraction, finding that the fit uncertainties obtained through the ``$\chi^2$
averaging'' method fully cover the range of $\alphasmZ$ variations driven by hadronization (finite
hadron-mass) corrections, different experimental final-hadron definitions, as well as overall fit procedure. 
An additional theoretical-scale uncertainty of +0.0015 has been calculated by redoing the
fit of the moments using a different shower cut-off value (i.e. relaxing the limiting-spectrum criterion).
The final value of $\alphasmZ$ obtained from a simple weighted-average of the couplings extracted independently
from the $\epem$ and DIS jet data, yields: $\alphasmZ$~=~0.1195$\,\pm\,$0.0021$\,$(exp)$\,^{+0.0015}_{-0.0}\,$(scale), 
which is in perfect accord with the current world-average of $\alphasmZ$~=~0.1185$\,\pm\,$0.0006 
(obtained at NNLO accuracy).
Work is in progress to combine the $\epem$ and DIS fragmentation-function moments into a {\it single} global
fit~\cite{DdE_RP}, which will better account for any residual correlated uncertainty. In
Fig.~\ref{fig:alphas_NLO} we compare our $\alphasmZ$ value to all other existing results at NLO accuracy. 
All the strong coupling values plotted are from the latest PDG compilation~\cite{PDG} except for 
the most recent jet results obtained from the CMS~\cite{CMS:2013yua} and ATLAS~\cite{Malaescu:2012ts} data.
Our result is the most precise of all approaches with
a totally different set of experimental and theoretical uncertainties. 
The weighted average of all the NLO values yields: $\alphasmZ$~=~0.1188$\,\pm\,$0.0012.
The methodology presented here provides a new promising approach for the determination of the QCD coupling
constant complementary to other existing jet-based methods --such as jet shapes, and inclusive 
jet production cross sections in $\epem$, DIS and p-p collisions-- and can be used to reduce the 
overall final uncertainty of the least well-known interaction coupling in nature.

\paragraph*{Acknowledgments}Redamy~P\'erez-Ramos acknowledges support from the Academy of Finland, Projects No. 130472 and
No.133005.


\end{document}